\documentclass[preprint,12pt]{elsarticle}

\usepackage{gensymb}

\usepackage{amssymb}
\usepackage{amsmath}

\usepackage{lineno}

\journal{NIM A}

\begin{document}

\begin{frontmatter}



\title{Properties of carbon-infused silicon LGAD devices after non-uniform irradiation with 24 GeV/c protons} 



\author[affl1]{C. Beir\~ao da Cruz e Silva\corref{cor1}}
\ead{cbeiraod@lip.pt}
\author[affl1]{G. Marozzo}
\author[affl1]{G. Da Molin}
\author[affl1]{J. Hollar}
\author[affl1]{M. Gallinaro}
\author[affl2]{M. Khakzad}
\author[affl2]{S. Bashiri Kahjoq}
\author[affl3]{K. Shchelina}

\cortext[cor1]{Corresponding author}
\affiliation[affl1]{organization={LIP Lisbon},
            addressline={Av. Prof. Gama Pinto, n.2}, 
            city={Lisboa},
            postcode={1649-003}, 
            country={Portugal}}

\affiliation[affl2]{organization={School of Particles and Accelerators, Institute for Research in Fundamental Sciences (IPM)},
            addressline={P.O. Box 19395-5531},
            city={Tehran},
            country={Iran}}

\affiliation[affl3]{organization={CERN},
  addressline={Esplanade des Particules 1},
            city={Geneva 23},
            postcode={CH-1121},
            country={Switzerland}}

\begin{abstract}
  Forward proton spectrometers at high-energy proton colliders rely on precision timing to discriminate signal from background. Silicon low gain avalanche diodes (LGADs) are a candidate
  for future timing detectors in these systems. A major challenge for the use of LGADs is that these detectors must be placed within a few mm of the beams, resulting in a very large
  and highly non-uniform radiation environment. We present a first measurement of the current and capacitance vs. voltage behavior of LGAD sensors, after a highly non-uniform irradiation
  with beams of 24 GeV/c protons at fluences up to $1\times10^{16} p/cm^{2}$.
\end{abstract}




\end{frontmatter}



\section{Introduction}
\label{sec1}

At the CERN Large Hadron Collider (LHC), forward proton detectors rely on precision timing measurements to distinguish signal events from "pileup" (multiple collisions in the same
bunch crossing) backgrounds. The CMS-TOTEM PPS and ATLAS-AFP projects pioneered precision proton timing in Run 2 of the LHC, using synthetic single-crystal
diamond~\cite{TOTEM:2017mgc,Bossini:2023uwa,Berretti:2016sfj,1826124,Bossini:2020pme,Bossini:2020ycc,DPNoteDiamonds} and quartz Cherenkov~\cite{Sykora:2020gzq, Cerny:2020nyj} technologies,
respectively. Initial studies were also performed with silicon low gain avalanche diode (LGAD) detectors~\cite{Arcidiacono:2017ibe} in 2017.

Since the early LGAD studies, the technology has matured, and is now the choice for the endcap timing layer upgrades of the CMS and ATLAS experiments~\cite{Butler:2019rpu, Torango, 2091129}.
This also makes them an attractive candidate for use in forward proton detectors at the High-Luminosity LHC (HL-LHC)~\cite{CMS:2021ncv}. A major challenge for using LGADs in forward proton
detectors is the radiation environment, which is both large and highly non-uniform, due to the proximity to the beam. While many previous studies have investigated the behavior of LGADs
after irradiation with reactor neutrons or proton beams~\cite{Ferrero:2018fen, Wu:2022ruu, Senger:2022zos, Li:2021hqx, Tan:2020usw, Padilla:2020sau, Mulargia:2024ikw}, these have focused
on the case of uniform irradiation, with a single device irradiated to a fixed amount. In this paper we aim to evaluate the behavior after highly non-uniform irradiation with high-energy
protons, to understand whether a single device can be operated after very different internal radiation doses.

\section{Radiation environment}
\label{radenv}

Forward proton detectors at the LHC are placed extremely close to the beam, at distances of a few mm. The sensor planes are oriented perpendicular to the beam, with one edge near the
beam and the other edge farther away. This results in a  highly non-uniform irradiation environment~\cite{CMS:2022hly,CMS:2021ncv}, with more than an order of magnitude variation over
distances of a few cm. At the High-Luminosity LHC, the sensors can be exposed to peak radiation doses of order $1 \times 10^{16} p/cm^{2}$ in the region closest to the beam. In the worst case, with no other mitigations, this dose can be reached in 1 year of HL-LHC running, which is the timescale for exchanging damaged sensors. 

\section{LGAD sensors}
\label{lgads}

\begin{figure}[h!]
  \begin{center}
\includegraphics[width=.5\textwidth]{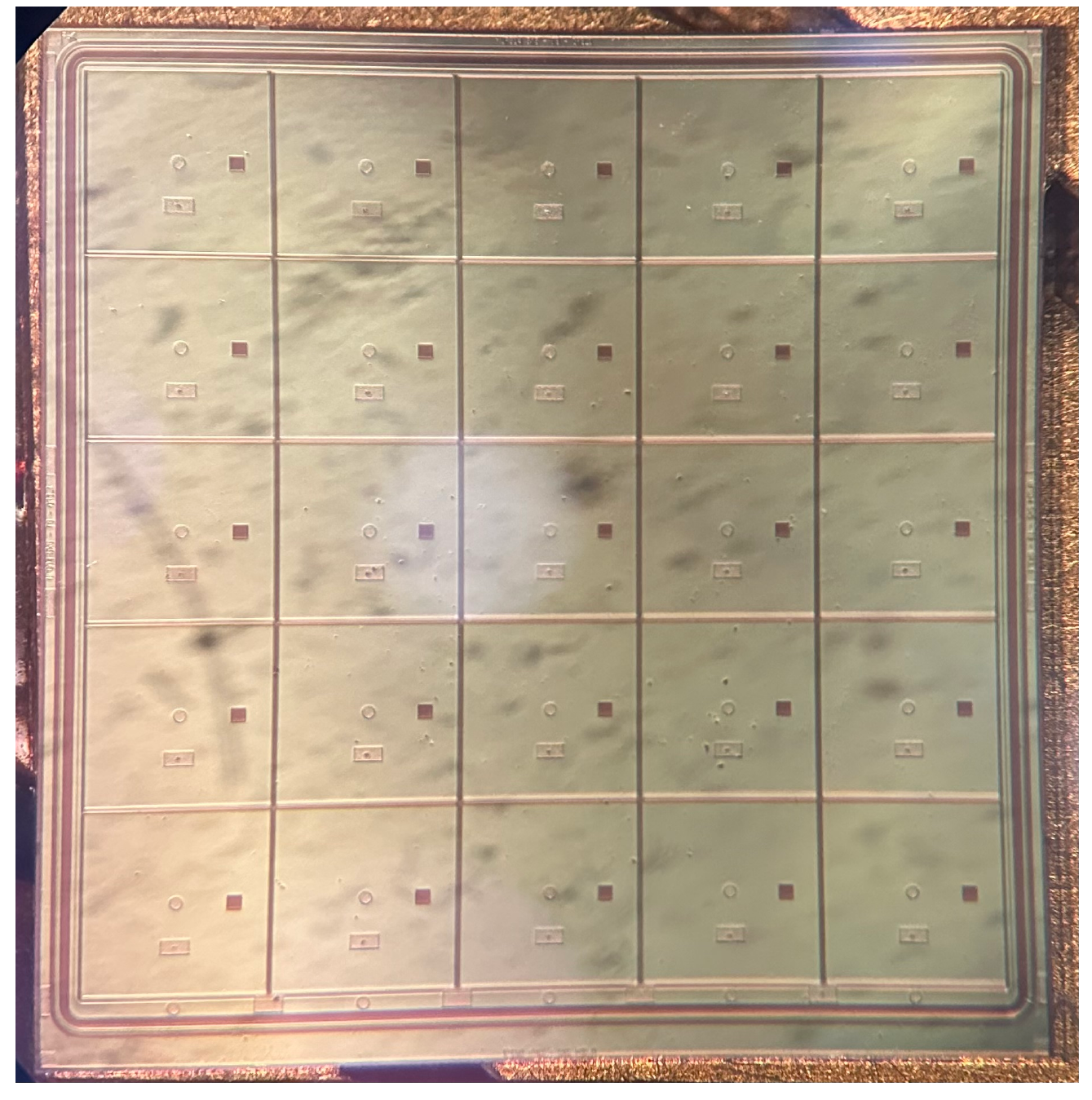}
  \end{center}
  \caption{One 5x5 LGAD used for this study, photographed under a microscope. The 5x5 square array of pixels is visible, inside the guard ring. The rectangular areas are the IV probe
    contact points for the pixels and the guard rings.}
  \label{fig:lgadmicroscope}
\end{figure}

The LGAD sensors used for this study were produced by Fondazione Bruno Kessler (FBK), as part of the UFSD4 production. They have pixels of area $1.3 \times 1.3  \text{mm}^2$, with a
total area defined by an array of 5x5 pixels. An example of a 5x5 LGAD used for this study is shown in Fig.~\ref{fig:lgadmicroscope}.

All of the sensors used in this study originate from wafer 18 of the FBK UFSD4 production, which was observed to be the most radiation hard of this production in uniform irradiation
tests. These sensors are designed with a deep gain layer, enriched with carbon to help mitigate the acceptor removal effect produced by radiation~\cite{Wu:2022ruu}. The properties of UFSD4 and 
wafer 18 are discussed in more detail in Ref.~\cite{Torango}.

In these sensors, a single bias voltage is provided to all pixels, as is typical of LGADs planned for the LHC experiments. This design imposes significant constraints on the HV
operating point. The HV must be large enough that the most irradiated regions of the detector still perform well, but small enough that the least irradiated regions do not go into breakdown.

\section{Irradiation conditions}
\label{irradcond}

\begin{figure}[h!]
  \begin{center}
\includegraphics[width=.45\textwidth]{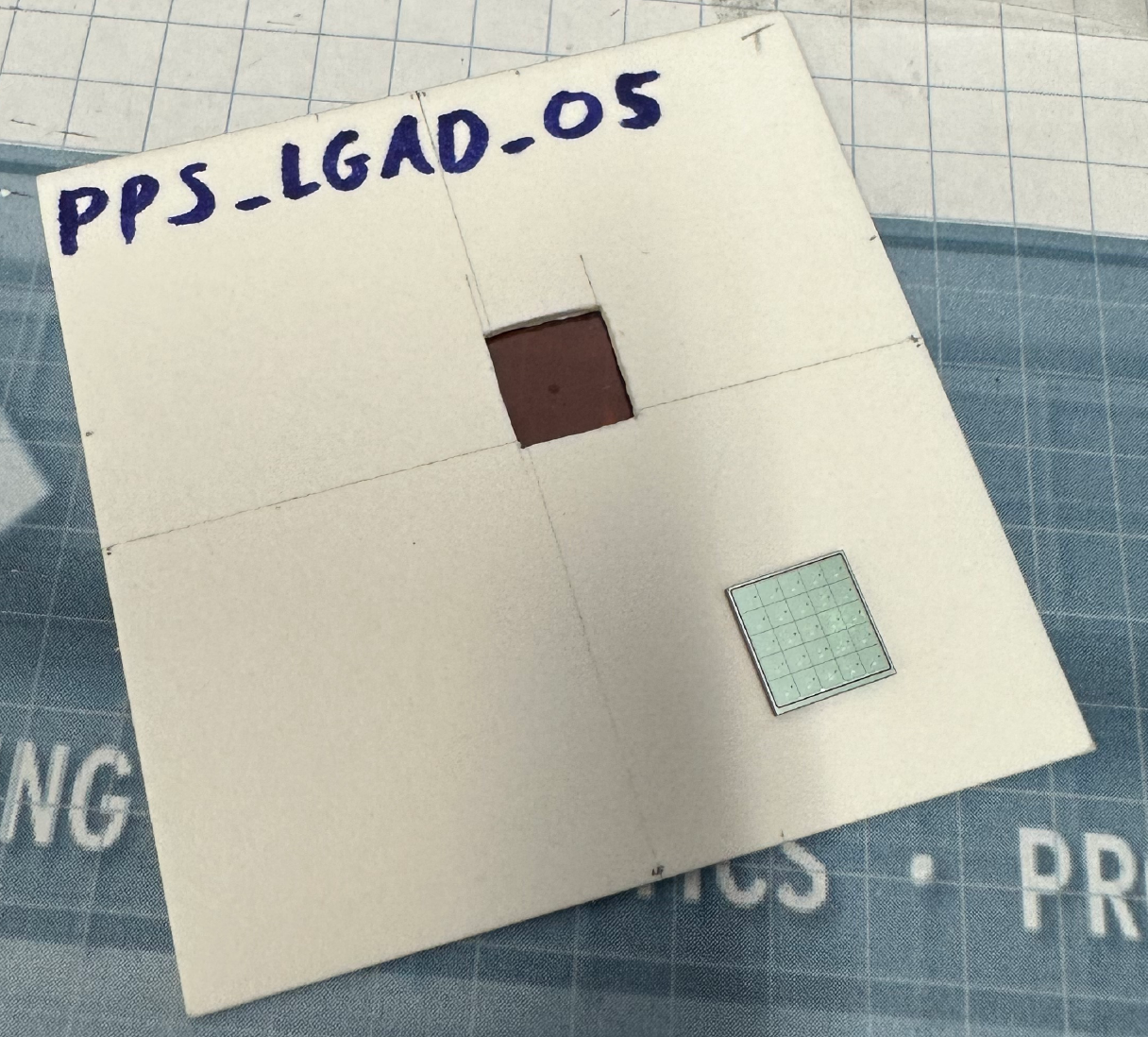}
 \includegraphics[width=.45\textwidth]{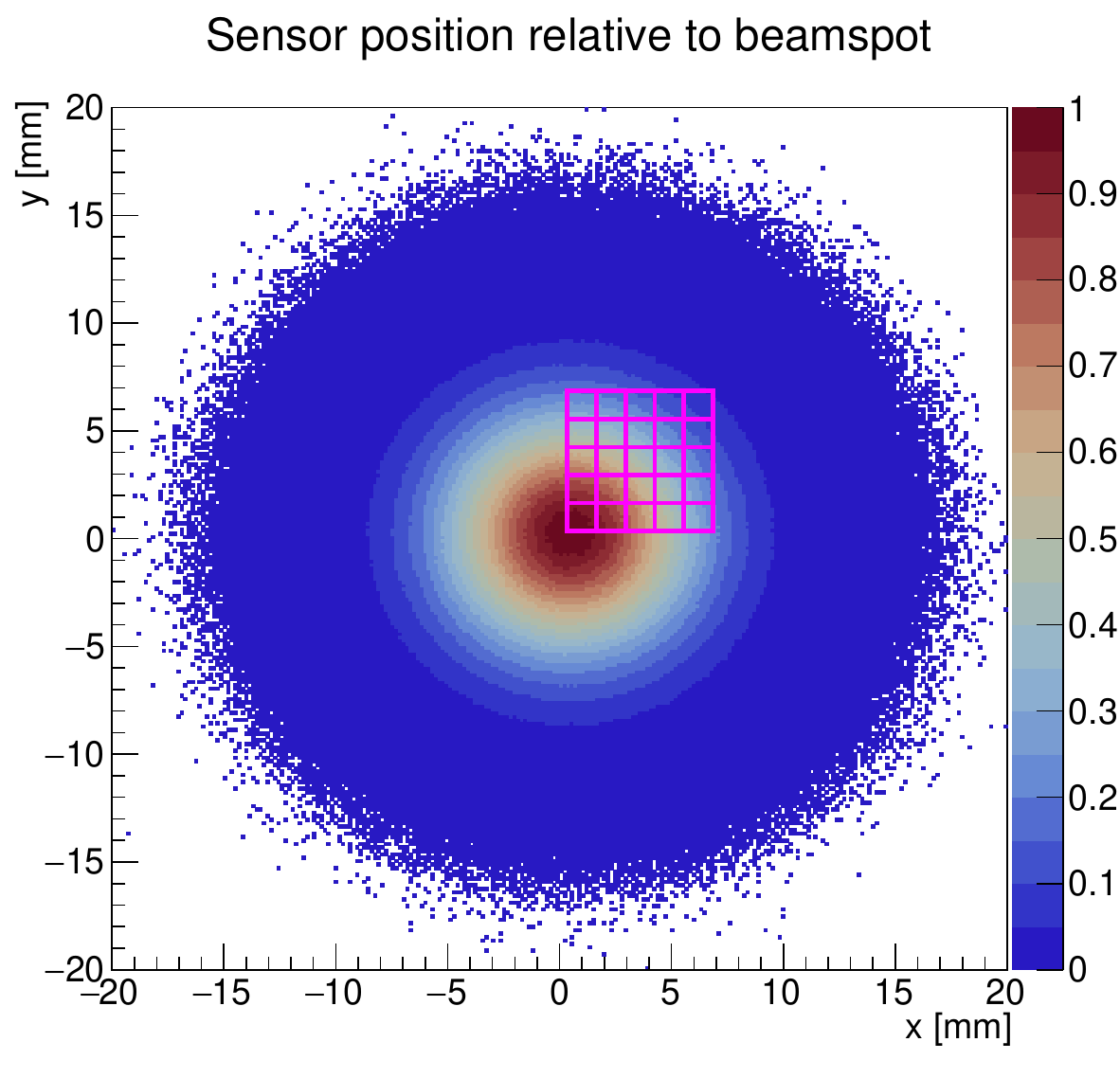}
 \end{center}
  \caption{Left: 5x5 LGAD sensor (seen before mounting at lower right) and frame used to mount the sensor for irradiation. The beam is centered at the center of the white square frame,
    with the sensor placed in the hole covered by kapton film. Right: Approximate positioning of sensors in the IRRAD beam. The squares indicate the positioning of the 5x5 LGAD device.
    The histogram indicates the beamspot shape during the period of irradiation, estimated from beam position monitoring data, with the colors indicating the relative intensity.}
  \label{fig:irradbeamspot}
\end{figure}

The irradiation was performed at the IRRAD facility~\cite{Ravotti:2014uda,Ravotti:2019dry} at the CERN Proton Synchrotron (PS). This provides high-intensity proton beams with a momentum
of 24 GeV/c. In order to produce a non-uniform gradient, the sensors were offset from the beam center in both the horizontal and vertical directions (Fig.~\ref{fig:irradbeamspot}),
creating a large variation in the dose along the diagonal going from the lower left to upper right of the sensor.

 \begin{figure}[h!]
  \begin{center}
 \includegraphics[width=.6\textwidth]{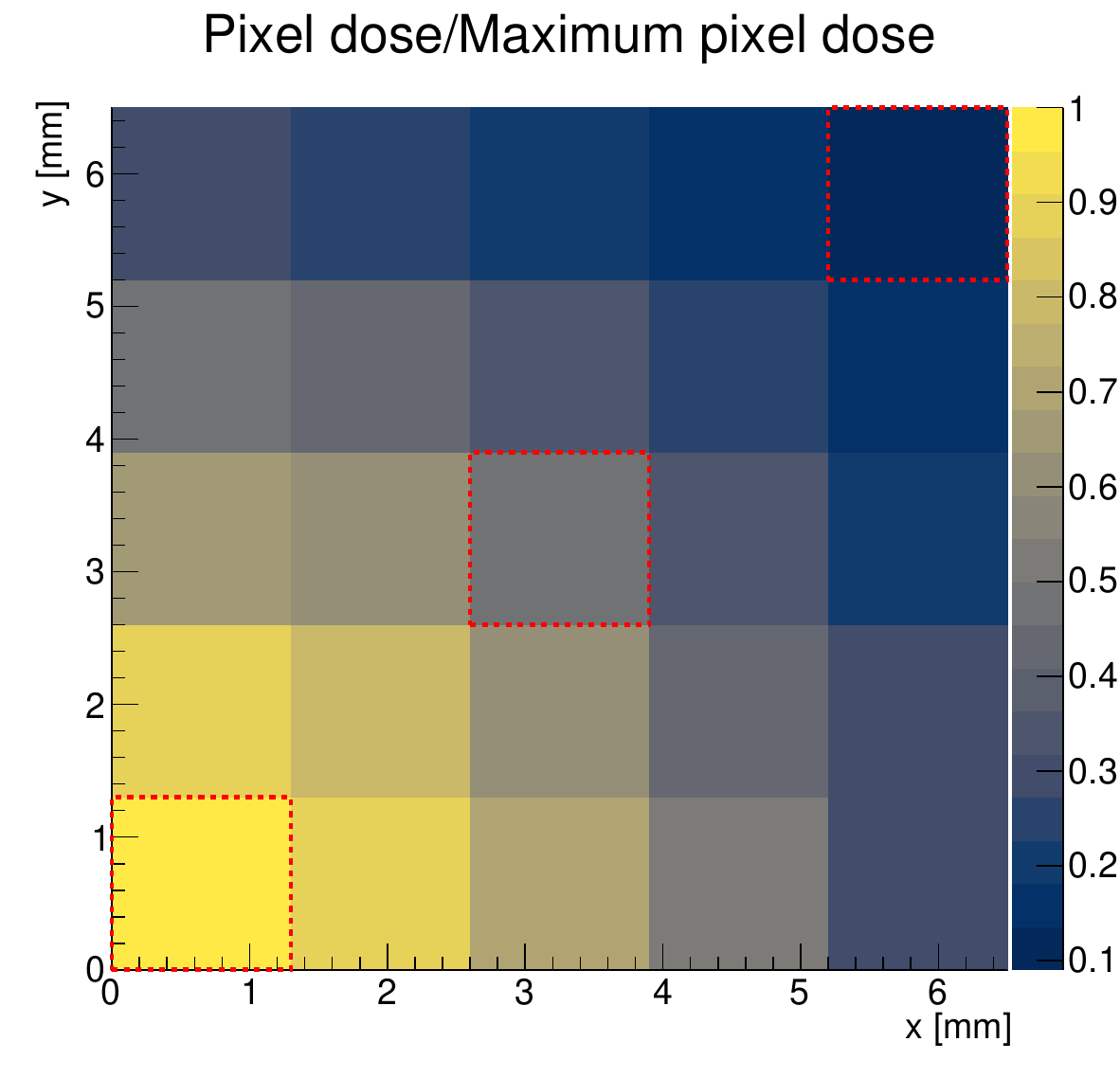}
  \end{center}
  \caption{Relative dose integrated in each pixel of the LGADs, compared to the most irradiated pixel. Each box in the plot represents a single $1.3 \times 1.3  \text{mm}^2$ pixel. The
    pixels for which IV curves are measured are shown by the dashed boxes at lower left, center, and upper right. The dimensions of the LGAD are approximate, and do not show the size of the
    guard ring or interpad spacing.}
  \label{fig:irradgradient}
\end{figure}

 A total of five sensors were tested. Two of the sensors were irradiated to reach a peak dose at the center of the beam corresponding to approximately $1 \times 10^{16}$ p/cm$^{2}$. Two
 other sensors were removed at the halfway point of the irradiation, resulting in a peak dose of approximately $5 \times 10^{15}$ p/cm$^{2}$. The remaining sensor was not irradiated, and
 served as a control sample. The actual dose was cross-checked using measurements of activated foils placed in the beamline. The dose measured at the center of the beam was found
 to be $3-11\%$ higher than the nominal target, with an uncertainty of 7\%.

 The irradiation gradient was estimated from beam position monitoring (BPM) measurements at the IRRAD facility. These measurements were used to generate a two-dimensional representation
 of the beamspot, and the corresponding radiation profile within the sensors under test. As shown in Fig.~\ref{fig:irradgradient}, there is approximately a factor of 2 in dose between
 the most irradiated pixel and the pixel at the center of the sensor, and a factor of 10 in dose between the most irradiated pixel and the least irradiated pixel.

\section{Measurement conditions}
\label{meascond}

The measurements of the IV and CV characteristics were conducted with a probe station at the CERN Solid State Detectors lab. A single probe was kept in contact with the guard ring and connected to ground,
while a second probe was used to measure the pixel of interest. The bias voltage was applied from the back side of the LGAD device, through the probe station chuck.

Prior to irradiation, the measurements were done both at room temperature, and at -20\degree C, consistent with the expected operating temperature of the LGADs. After irradiation, the
measurements were performed only at -20\degree C, and the samples were also stored at -20\degree C to limit annealing effects. In the following we focus only on the measurements
performed at -20\degree C, in order to compare the results before and after irradiation in the same conditions.

The measurements were done by increasing the voltage from zero to the maximum, and then decreasing from the maximum to zero to check for hysteresis effects. The maximum voltage was
chosen to be in a safe range such that the sensors did not approach breakdown. This maximum was between 200V and 400V, depending on the sensor and radiation dose.

\section{Results}
\label{results}

The LGAD IV curves were measured before and after irradiation. The focus was on pixels positioned on the diagonal of the LGAD sensor, which followed the direction of the irradiation
gradient. Specifically pixels [0,0], [2,2], and [4,4] were measured, where [0,0] is located at the bottom left and [4,4] at the upper right of the sensor shown in
Figure~\ref{fig:irradgradient}. In this way the most-irradiated and least-irradiated pixels were tested, as well as a pixel with an intermediate radiation dose.

\subsection{IV measurements before and after irradiation}
\label{subseciv}

\begin{figure}[h!]
  \begin{center}
\includegraphics[width=.45\textwidth]{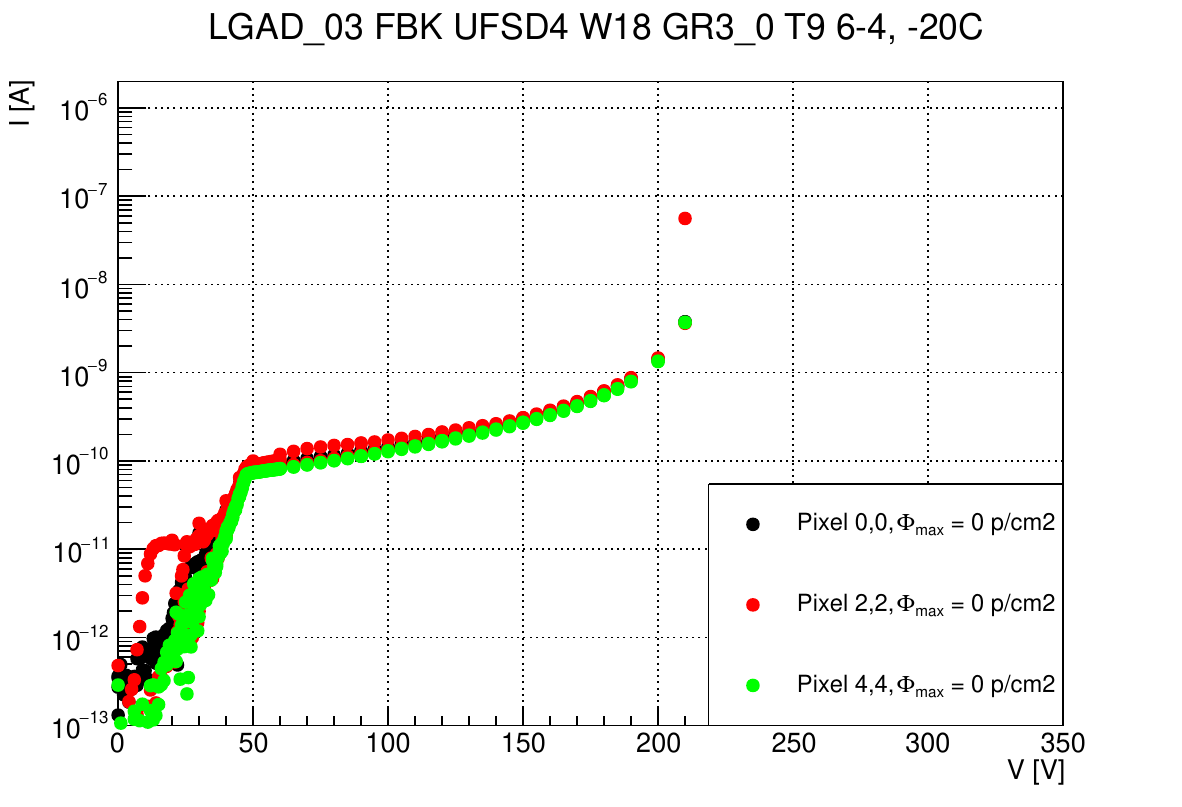}
  \end{center}
  \caption{IV curves for the non-irradiated sensor, measured for three pixels along the diagonal. The black, red, and green points represent the pixels along the diagonal in
    positions [0,0], [2,2], and [4,4], respectively.}
  \label{fig:ivcurves0}
\end{figure}

\begin{figure}[h!]
  \begin{center}
\includegraphics[width=.45\textwidth]{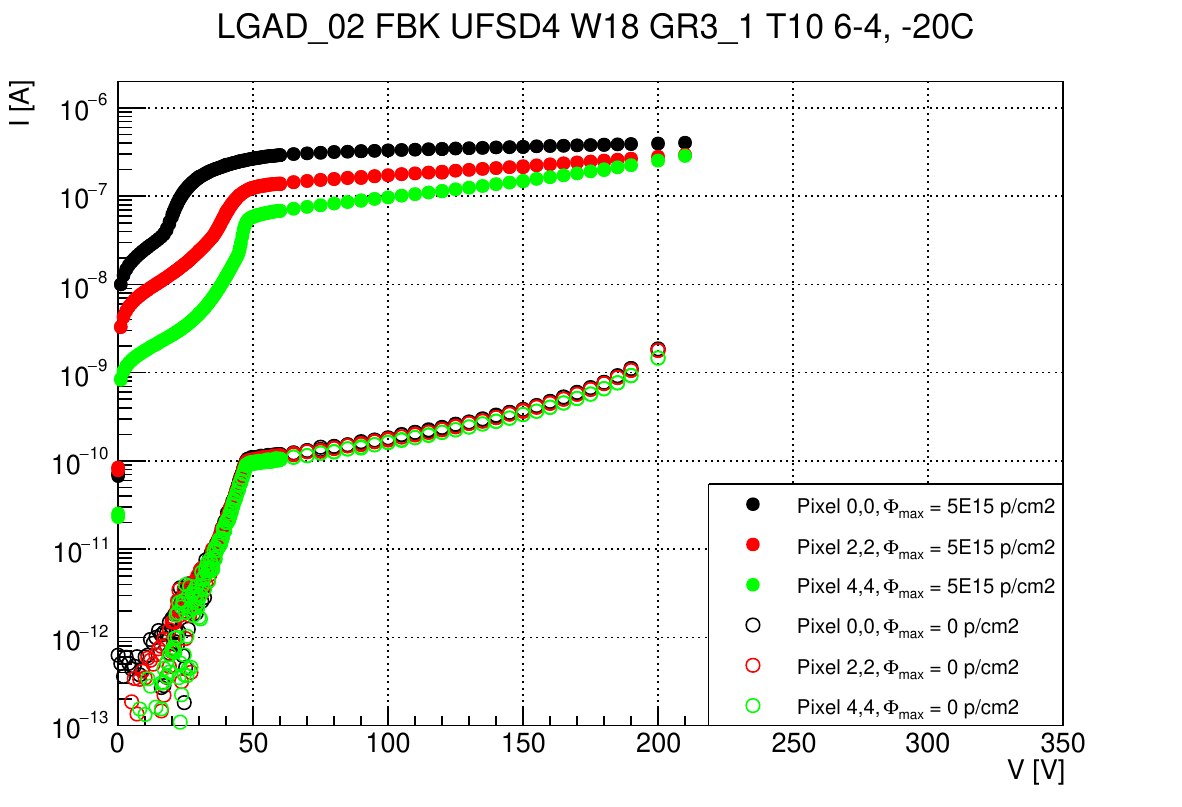}
\includegraphics[width=.45\textwidth]{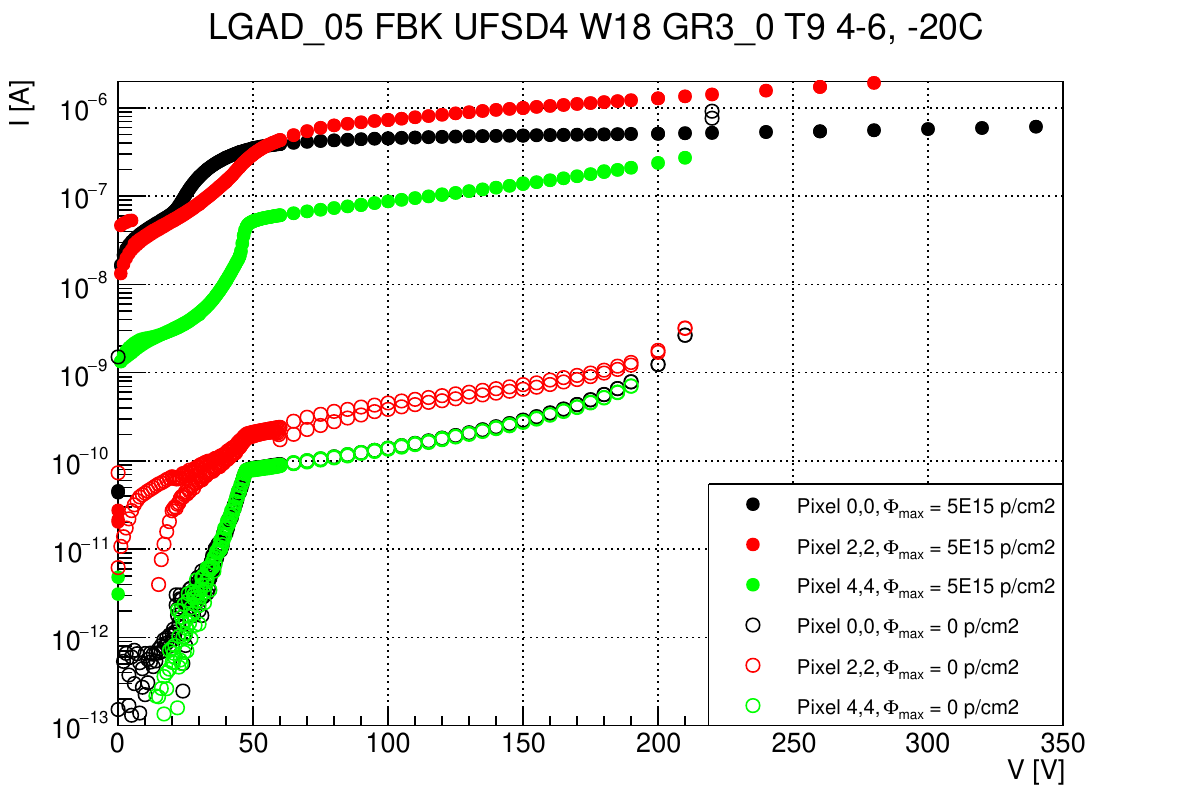}
  \end{center}
  \caption{IV curves before and after irradiation, measured for three pixels along the direction of the irradiation gradient, for two devices irradiated to a maximum
    of $5\times10^{15} p/cm^{2}$. The open markers represent the measurement before irradiation, the full markers the measurement after irradiation. The black, green, and red points
    represent the most irradiated, least irradiated, and an intermediate pixel, respectively.}
  \label{fig:ivcurves5e15}
\end{figure}

\begin{figure}[h!]
  \begin{center}
\includegraphics[width=.45\textwidth]{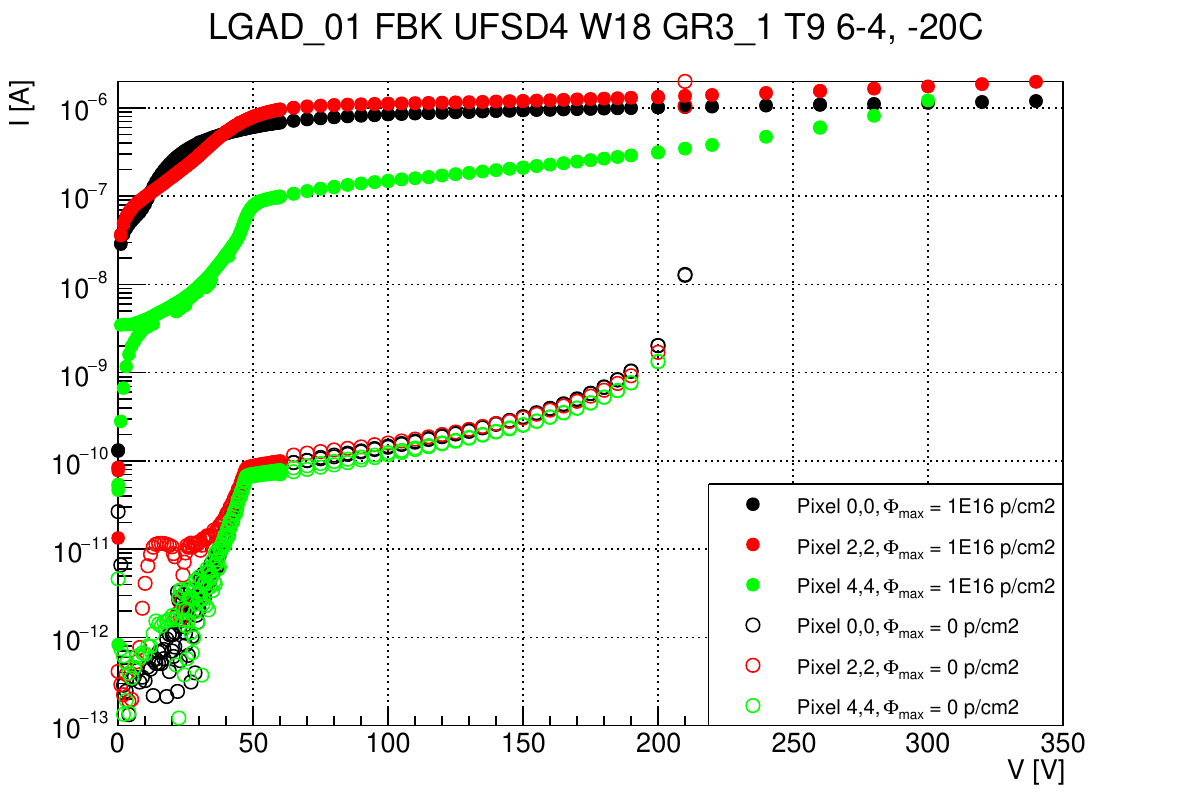}
\includegraphics[width=.45\textwidth]{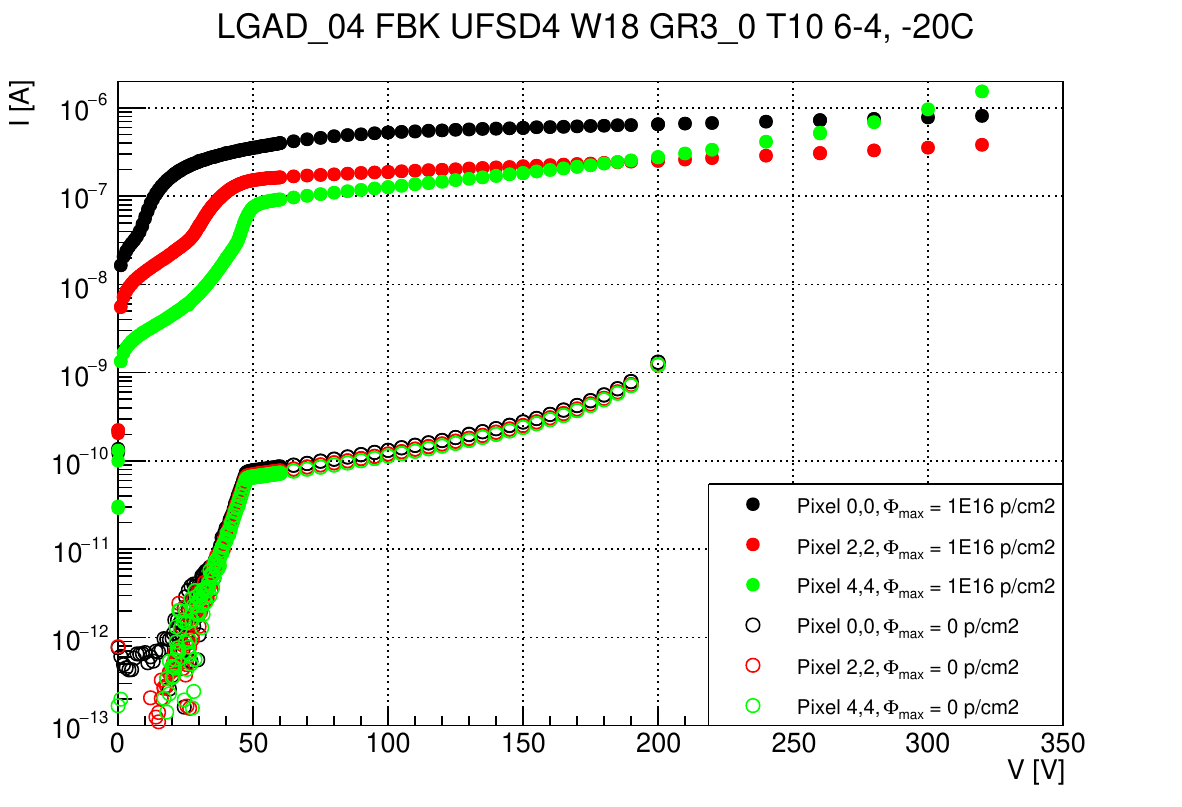}
  \end{center}
  \caption{IV curves before and after irradiation, measured for three pixels along the direction of the irradiation gradient, for two devices irradiated to a maximum
    of $1\times10^{16} p/cm^{2}$. The open markers represent the measurement before irradiation, the full markers the measurement after irradiation. The black, green, and red points
    represent the most irradiated, least irradiated, and an intermediate pixel, respectively.}
  \label{fig:ivcurves1e16}
\end{figure}

Fig.~\ref{fig:ivcurves0} shows the current as a function of voltage for the sensor that was not irradiated. Figures~\ref{fig:ivcurves5e15} and ~\ref{fig:ivcurves1e16} show the IV curves
for three pixels on each of the four devices under test, before and after irradiation.  All of the plotted measurements were performed with the sensors cooled to -20\degree C.

Prior to irradiation, all pixels measured show generally similar behavior. A steep increase is observed, reaching a sharp "knee" around 50~V, as the sensor reaches full depletion
voltage. This is followed by a  plateau-like behavior with a slow increase of currents up to  around 0.1 nA. At higher voltages, at approximately 200~V, the current shows a runaway increase
as it approaches the breakdown voltage.

After irradiation, the effect of the non-uniformity is clear, as the least irradiated pixel maintains the sharpest turn-on. The more irradiated pixels display a shallower turn-on, and
operating voltages with currents of order 1 $\mu A$. The most irradiated pixels also do not show any sign of a rapid increase toward breakdown, up to the highest values measured (350 V).
In all cases the currents at the operating voltages after irradiation are 3-4 orders of magnitude larger than pre-irradiation.

\subsection{Operating voltage ranges}
\label{subsecoprange}

Operating the sensors with a single high voltage setting under non-uniform irradiation requires a working point where the most irradiated areas of the sensor are above the full depletion
voltage, while the least irradiated areas of the sensor are safely below breakdown.

\begin{figure}[h!]
  \begin{center}
\includegraphics[width=.4\textwidth]{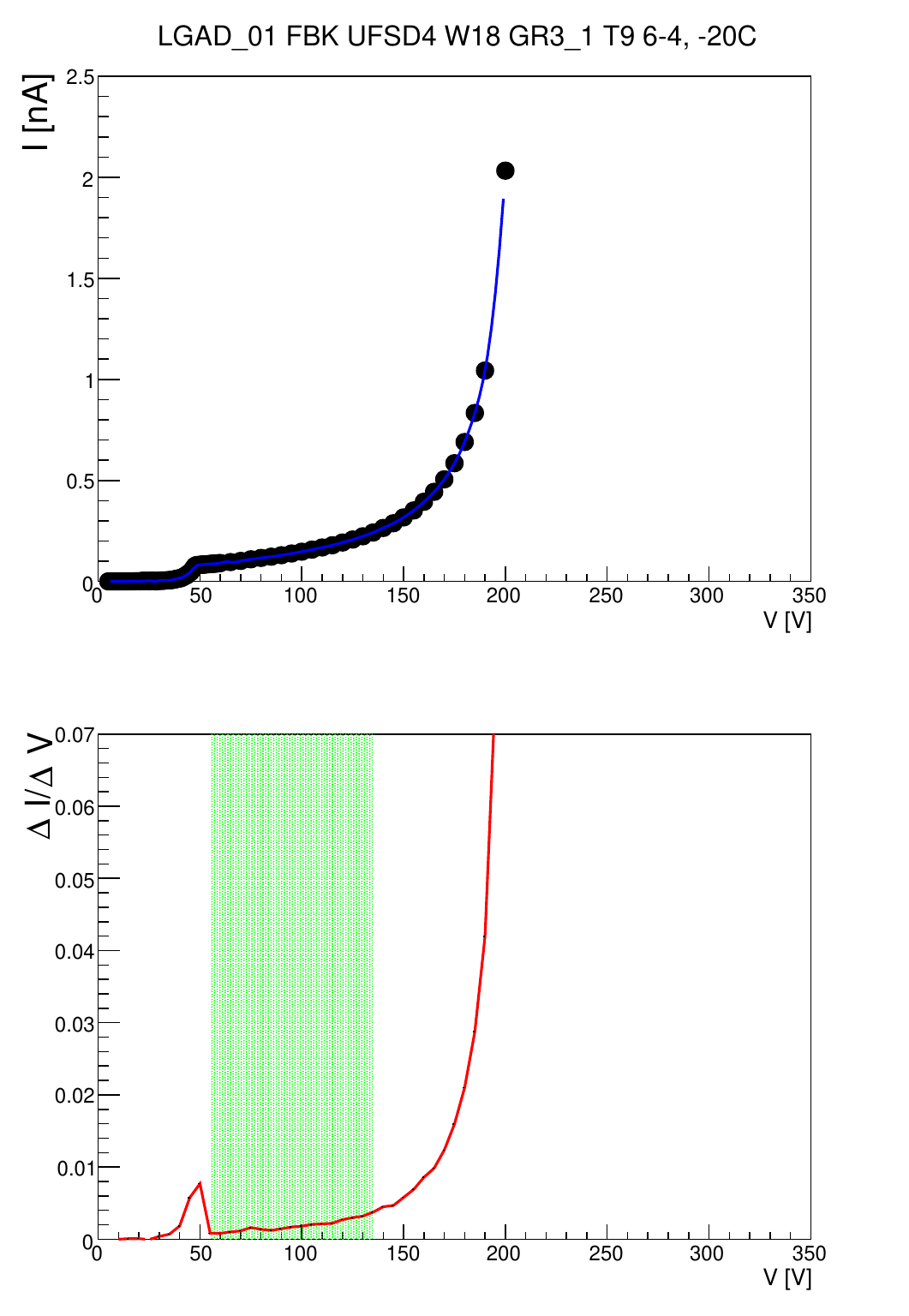}
\includegraphics[width=.4\textwidth]{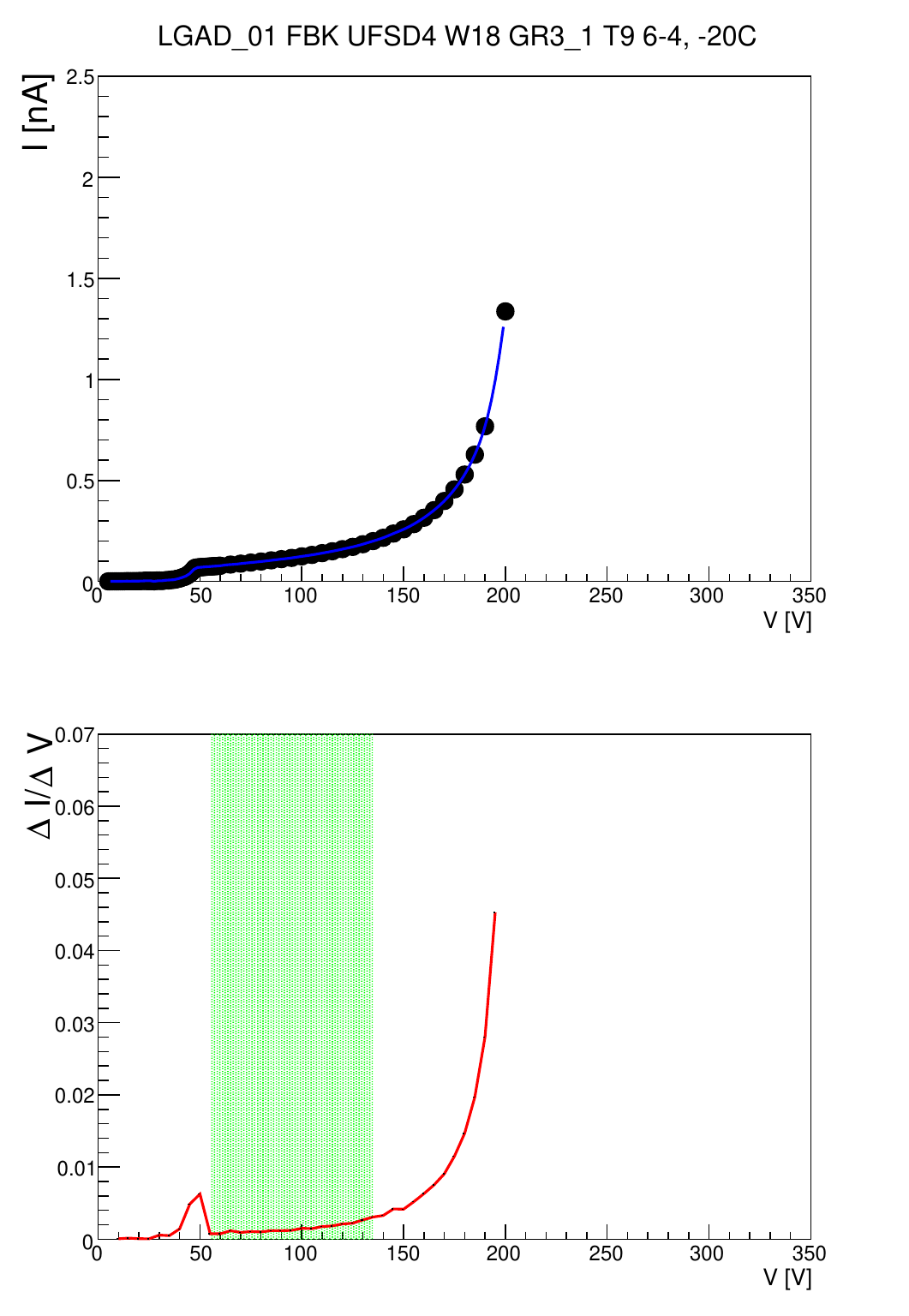}
  \end{center}
  \caption{IV curves for pixels 0,0 (left) and 4,4 (right) of device 1, before   irradiation. The measured points and spline interpolation are shown above, and the derivative of the
    interpolation is shown below. The shaded region in the lower plots indicates the operational voltage range.}
  \label{fig:ivcurvefitpreirrad}
\end{figure}

\begin{figure}[h!]
  \begin{center}
\includegraphics[width=.4\textwidth]{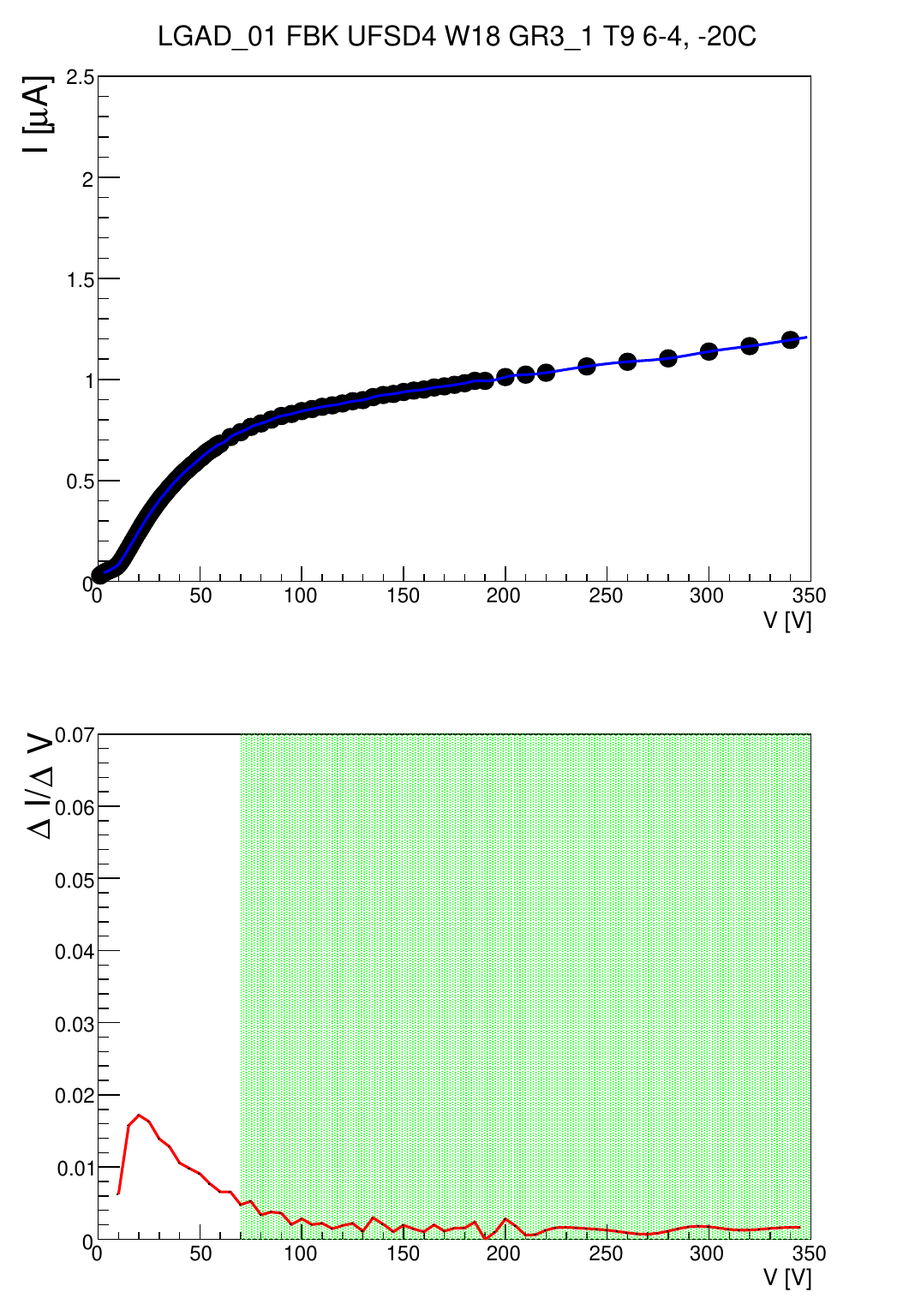}
\includegraphics[width=.4\textwidth]{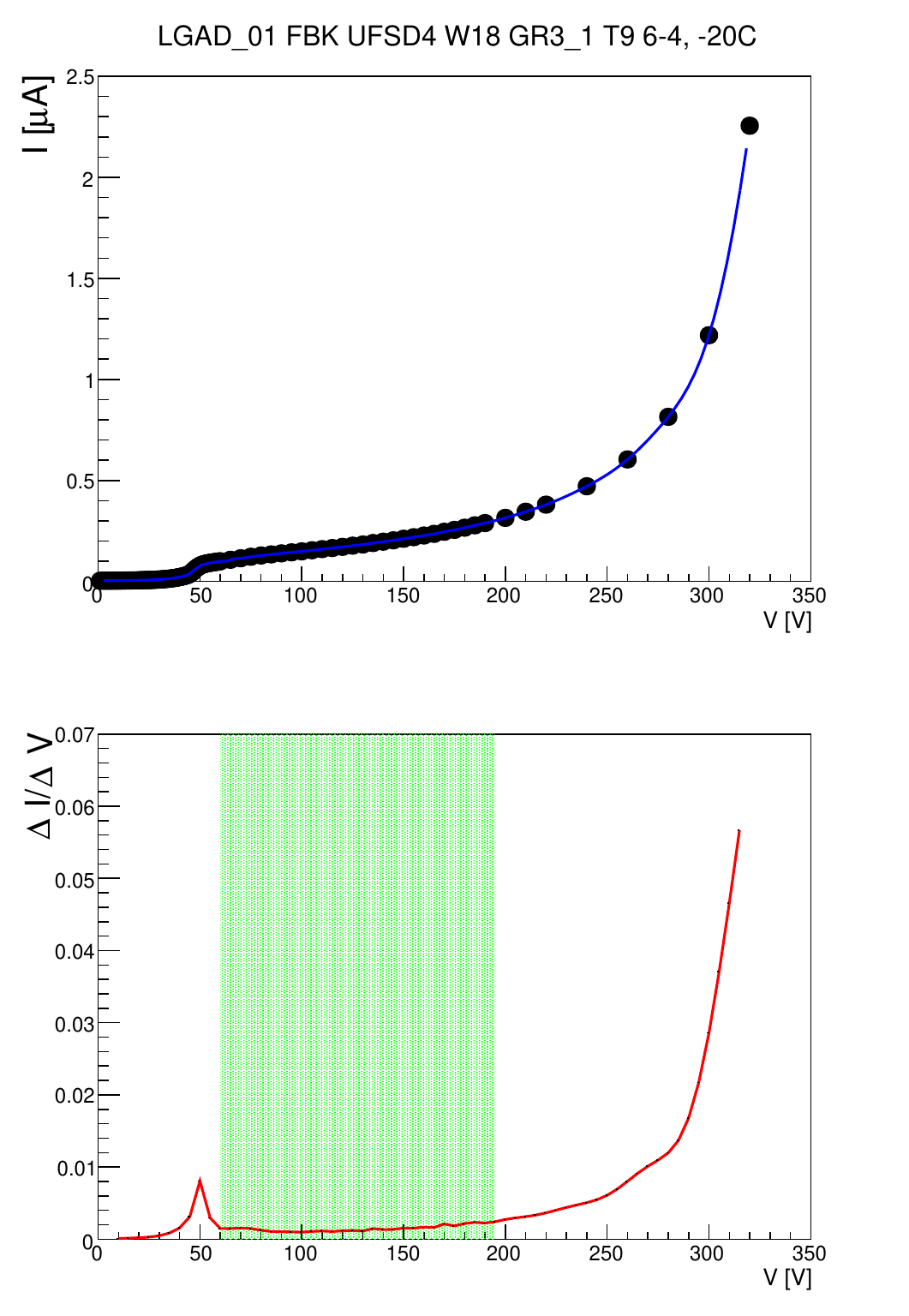}
  \end{center}
  \caption{IV curves for pixels 0,0 (left) and 4,4 (right) of device 1, after irradiation with a maximum dose of $1 \times 10^{16} p/cm^{2}$. Pixel 0,0 is near the beam center and
    receives the maximum dose, while Pixel 4,4 is farthest from the beam center. The measured points and spline interpolation are shown above, and the derivative of the interpolation
    is shown below. The shaded region in the lower plots indicates the operational voltage range.}
  \label{fig:ivcurvefitpostirrad}
\end{figure}

In order to quantify the operating voltage range, we analyze the shape of the IV curves. A cubic spline interpolation is performed between consecutive voltage points, using 5 V steps to
smooth fluctuations. The derivative dI/dV is then  calculated based on the interpolation. The derivative has a sharp peak corresponding to the sharp rise of current as the bulk becomes
activated, followed by a plateau-like behavior above the knee. For the less irradiated pixels, the derivative begins increasing again at higher voltages as the sensor approaches the
breakdown region. For the most irradiated pixels, the derivative remains roughly constant up to the highest voltages tested. We consider the range between the initial peak in the derivative
and the second increase at larger voltages as the operating range for each pixel. Quantitatively, we define the maximum operating voltage as the point at which $dI/dV$ reaches $50\%$
of the initial peak. The procedure is illustrated in Figures~\ref{fig:ivcurvefitpreirrad} and~\ref{fig:ivcurvefitpostirrad}
for two pixels of one LGAD device, before and after irradiation.

In Table~\ref{tab:vopranges}, the upper and lower operating voltages obtained are listed for each pixel under test, before and after irradiation. In cases where no increase of the
derivative corresponding to breakdown is observed, the maximum voltage tested is shown as a lower limit.

\begin{table}[ht!]
\begin{center}
\begin{tabular}{l|l|c|c|c}\hline
Peak irradiation & Device & Pixel & $V_{OP} [V]$ & $V_{OP} [V]$ \\ (p/cm$^{2}$) & & & pre-irradiation & post-irradiation \\\hline
 & & 0,0 & 55 - 135 & 70 - $>$400  \\
$1\times10^{16}$ & 1 & 2,2 & 55 - 145 & 70 - $>$400 \\
 & & 4,4 & 55 - 135 & 60 - 195 \\\hline
& & 0,0 & 55 - 135 & 65 - $>$320 \\
$1\times10^{16}$ & 4 & 2,2 & 55 - 135 & 55 - $>$320 \\
 & & 4,4 & 55 - 140 & 55 - 195 \\\hline
& & 0,0 & 55 - 145 & 50 - $>$200 \\
$5\times10^{15}$ & 2 & 2,2 & 55 - 135 & 55 - $>$200\\
 & & 4,4 & 55 - 140 & 55 - 185 \\\hline
& & 0,0 & 55 - 135 & 60 - $>$400 \\
$5\times10^{15}$ & 5 & 2,2 & 55 - 110 & 85 - 200 \\
 & & 4,4 & 55 - 135 & 55 - 180 \\\hline
& & 0,0 & 55 - 135 & N/A \\
0 & 3 & 2,2 & 55 - 140 & N/A \\
 & & 4,4 & 55 - 135 & N/A \\\hline

\end{tabular}
\caption{Operating voltage ranges of LGAD sensors before and after irradiation, based on IV curve shapes. For each sensor, the minimum and maximum voltages are shown for 3 pixels along
  the irradiation gradient, before and after irradiation. The ranges are determined by analyzing the derivative of the interpolation of the measured data points for each pixel.}
\label{tab:vopranges}
\end{center}
\end{table}

Prior to irradiation, this procedure finds a minimum operating voltage for almost all pixels close to 55~V. The maximum, determined by the point where the slope of the IV curve starts
rapidly increasing, is in most cases between 135~V and 145~V. After irradiation, the minimum voltages range from 50~V to 85~V, depending on the sensor and radiation dose received. For the
least irradiated pixels, the maximum voltages are between 180~V and 200~V, while for the most irradiated pixels no maximum is found up to the largest voltages measured (between 200~V and
400~V, depending on the device).

For all devices tested, the maximum $V_{OP}$ of the least irradiated pixel is above the minimum $V_{OP}$ of the most irradiated pixel,
indicating that a common operating voltage range is possible with an order of magnitude gradient in the radiation dose. The
minimum $V_{OP}$ of the most irradiated pixels is also below the maximum $V_{OP}$ of the non-irradiated pixels. We note that in some
cases the operating voltage of pixel [2,2] is as large or larger than the other pixels with a somewhat different trend, even
before irradiation. This is believed to be due to the lack of grounding on adjacent pixels, where the centermost pixel is
most susceptible to leakage from neighboring pixels.

\subsection{Gain Layer Voltage and Acceptor Removal}
\label{subsecglremove}

The mechanism of radiation damage on LGADs is believed to be at least partly related to the deactivation of boron doping atoms, resulting in an effective degradation or removal of acceptors.

In order to study the acceptor removal from the IV curves, we use the k-factor method~\cite{Bacchetta:2001rf}. This is similar to the dI/dV derivative method used to define the operating
voltage range, except the derivative is multiplied by the ratio of current to voltage to obtain the dimensionless quantity:

$$k = dI/dV * V/I$$

\begin{figure}[h!]
  \begin{center}
\includegraphics[width=1\textwidth]{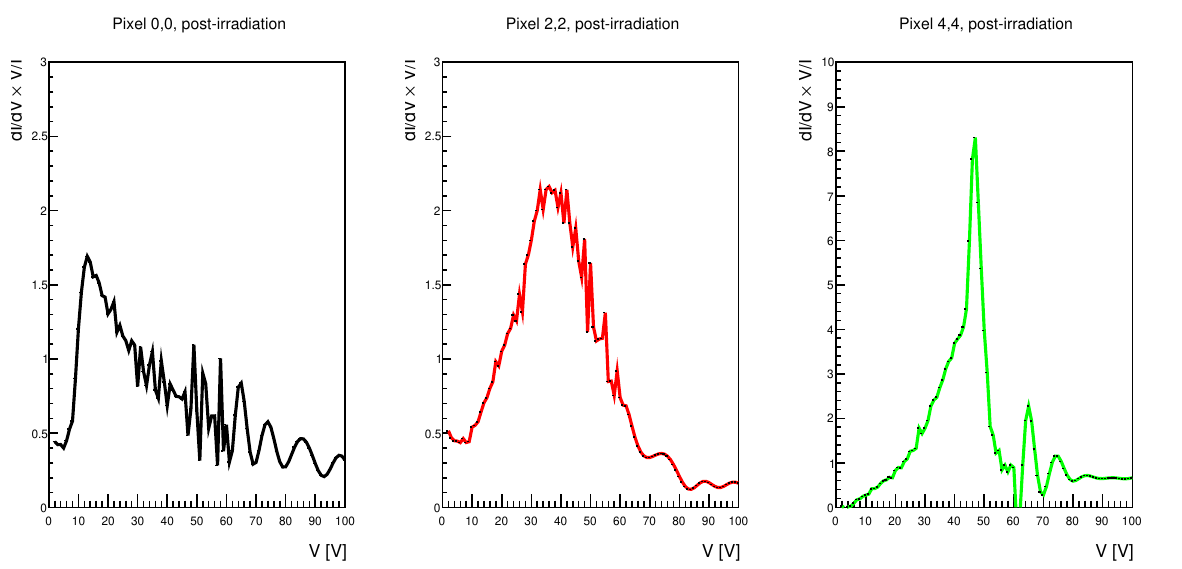}
  \end{center}
  \caption{K-factor distributions of one LGAD irradiated to a maximum dose of $1 \times 10^{16}$p/cm$^{2}$. The distributions are shown for the most (left) and least (right)
    irradiated pixels, and for an intermediate (center) pixel.}
  \label{fig:kfit}
\end{figure}

The first maximum of this function occurs at decreasing voltages, as the irradiation dose increases and a larger fraction of acceptors are removed. 
The maximum of the k-factor below breakdown is used to define the gain layer voltage, $V_{GL}$. The acceptor removal function is then defined by the ratio  $V_{GL}(\Phi)/V_{GL}(0)$,
where $\Phi$ represents the radiation dose, $V_{GL}(\Phi)$ is the gain layer voltage at a given radiation dose, and $V_{GL}(0)$ is the gain layer voltage prior to irradiation.
For the $V_{GL}(0)$ measurement before irradiation, the currents below the sharp increase are very low, near the resolution of the measurement device. In order to avoid
large fluctuations, a smoothing over 5~V steps is used, as in the case of the operating voltage range study. After irradiation, the currents are much larger, so a 1~V granularity
of the measurement is used. These 5~V and 1~V ranges are propagated into the ratio as uncertainties in the acceptor removal fraction. In addition, the $7\%$ uncertainty from the IRRAD
foil activation measurements is included as an uncertainty on the accumulated radiation dose. The k-factor distributions for one device after irradiation are shown
in Fig.~\ref{fig:kfit}.

\begin{figure}[h!]
  \begin{center}
\includegraphics[width=.8\textwidth]{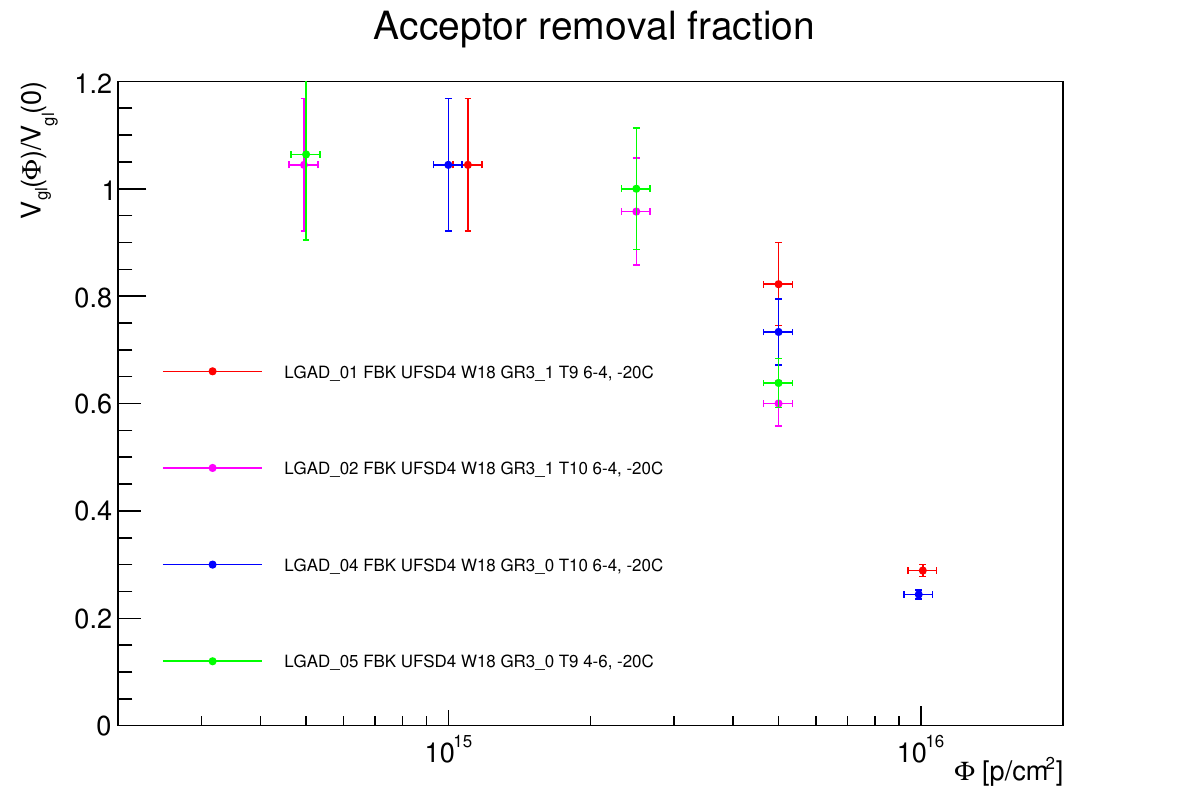}
  \end{center}
  \caption{Acceptor removal determined from IV curves, for pixels exposed to different radiation levels. The two sensors exposed to a maximum of $1 \times 10^{16}$p/cm$^{2}$ are shown by\
    the red and blue points. The two sensors exposed to a maximum of $5 \times 10^{15}$p/cm$^{2}$ are represented by the purple and green points. The vertical error bars represent the
  voltage step size used in the IV curve scan. The horizontal error bars represent the uncertainty in the accumulated radiation dose, obtained from foil activation measurements.}
  \label{fig:removal}
\end{figure}

Figure~\ref{fig:removal} shows the calculated acceptor removal fraction, as a function of the approximate radiation dose accumulated on each pixel. The approximate dose on each pixel, plotted on the x axis, is estimated from the beamspot profiles (Fig.~\ref{fig:irradgradient}). The measured pixels reveal a generally 
consistent acceptor removal fraction as a function of the irradiated dose, despite belonging to different sensors. We note that at $5 \times 10^{15}$p/cm$^{2}$, the values obtained from pixel [2,2] of the more irradiated devices tend to be higher than those from pixel [0,0] of the less irradiated devices, despite receiving the same integrated dose. This may be consistent with a greater effect of leakage from the ungrounded neighboring pixels in the central pixels. Even when the central pixels are excluded, the sensors show a remaining acceptor fraction at the highest doses that is larger than some previous measurements with other devices~\cite{Moll:2020kwo,Padilla:2020sau,Gkougkousis:2021ofs}. This may be due to choosing carbon-implanted sensors from the wafer that was observed to be the most radiation-hard of this production. Further measurements should be done on similar sensors to confirm this result. 

\subsection{CV measurements before and after irradiation}
\label{subseccv}

For one of the sensors irradiated to the maximum peak dose of 1$\times 10^{16}$p/cm$^{2}$, capacitance versus voltage measurements were made before and after irradiation, at -20\degree C.
The capacitance was measured up to voltages of 65~V, again with both increasing and decreasing voltages to check for hysteresis effects. The same measurements were also performed for
the non-irradiated sensor.

\begin{figure}[h!]
  \begin{center}
\includegraphics[width=.45\textwidth]{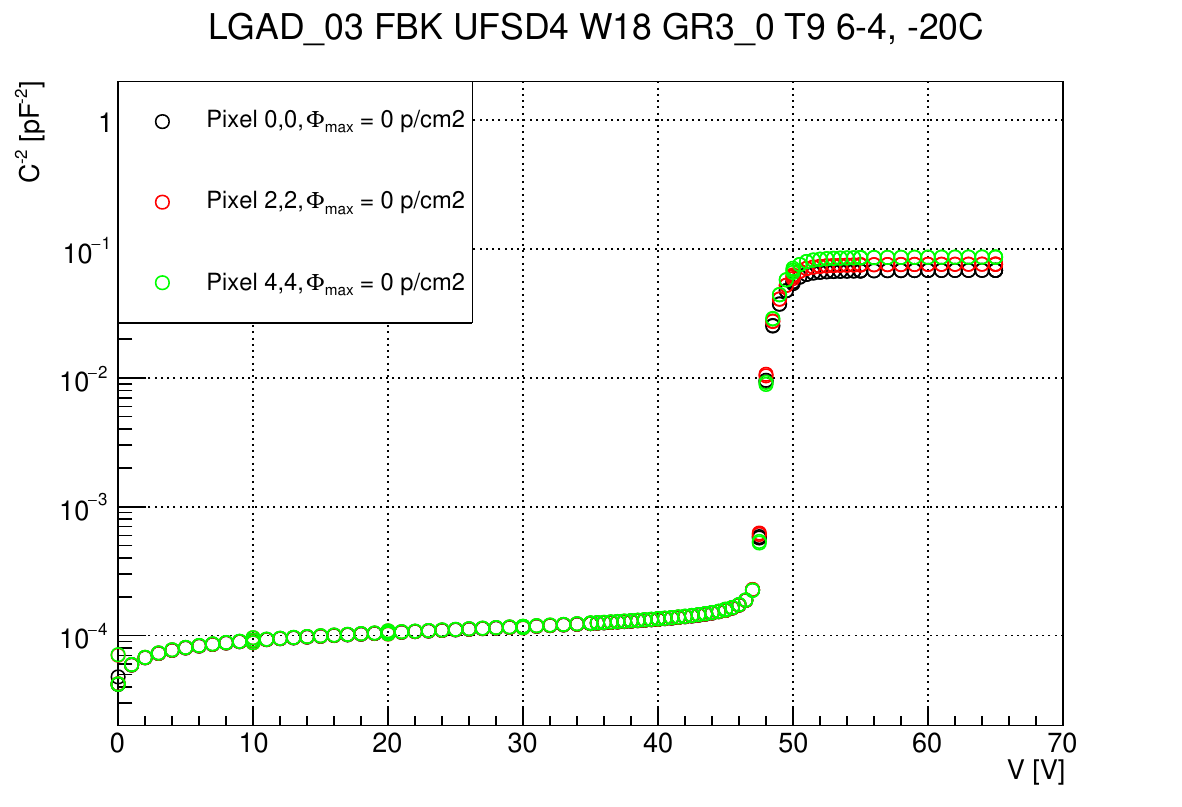}
\includegraphics[width=.45\textwidth]{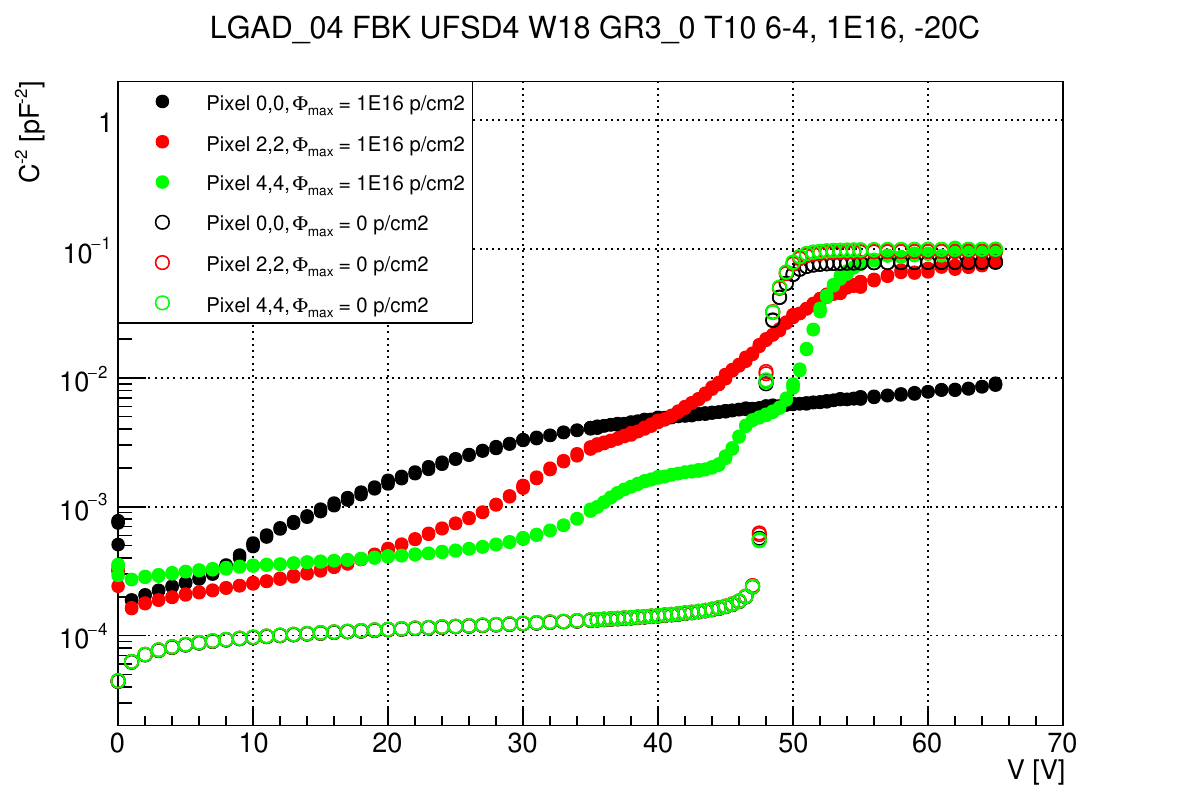}
  \end{center}
  \caption{CV curves measured for two sensors at -20\degree C. Left: curves for the non-irradiated sensor, measured for three pixels along the diagonal. Right: curves before and after
    irradiation, measured for three pixels along the direction of the irradiation gradient, for one device irradiated to a maximum of $1\times10^{16} \text{p/cm}^{2}$. The open markers
    represent the measurement before irradiation, the full markers the measurement after irradiation.}
  \label{fig:cvcurves4}
\end{figure}

The results are shown in Fig.~\ref{fig:cvcurves4}, plotted as $1/C^{2}$ vs. voltage. For the non-irradiated sensor, all pixels show a similar behavior, with a nearly flat curve as the
gain layer is activated, followed by a sharp turn-on around 45 V as the bulk becomes depleted, followed by a plateau around 50 V after the sensor reaches full depletion. This value is
compatible with the minimum operating voltages obtained from the analysis of the IV curves before irradiation. The other sensor that was irradiated to the maximum fluence shows very similar
behavior prior to irradiation.

After irradiation, the voltage at which the gain layer is depleted is observed to decrease for the more irradiated pixels, while the slope during the depletion of the bulk becomes
shallower. The capacitance at the lowest voltages is observed to decrease for all pixels after irradiation, consistent with a decreasing depth of the gain layer. This behavior within a single
device is qualitatively similar to Refs.~\cite{Padilla:2020sau,Wu:2022ruu}, where different devices were tested with different uniform radiation doses. For the most irradiated pixel, the
capacitance does not reach a clear plateau up to the maximum voltage of 65 V. For the less irradiated pixels, receiving doses of $\sim$5$\times 10^{15}$p/cm$^{2}$
and $\sim$1$\times 10^{15}$p/cm$^{2}$, the minimum capacitance is reached by 65 V, slightly higher than the value seen before irradiation.

\section{Conclusions}
\label{conclusions}

The use of silicon LGAD timing detectors for forward protons at the HL-LHC requires operation in a new radiation environment. Both the peak irradiation and the size of the non-uniform
irradiation gradient are larger than expected for other applications, even at the HL-LHC.

A first measurement of carbon-infused LGAD current and capacitance vs. voltage properties was performed before and after non-uniform irradiation with 24 GeV/c proton beams at the CERN PS, with a
peak dose of up to 1$\times 10^{16}$p/cm$^{2}$. A gradient of a factor $\sim$10, realistic for forward proton detectors, was achieved by offsetting the sensors from
the beam center in the vertical and horizontal directions. After irradiation, all pixels were found to reach an operational voltage at or below 90~V, with currents of order $\sim$1$~\mu A$ at
a temperature of -20\degree C. The less irradiated pixels were observed to approach breakdown above 200 V. This indicates that a common high voltage working point may be found for all pixels in a
sensor, in spite of the large differences in radiation dose.

Measurements of the CV curves for one sensor show that pixels irradiated up to approximately $5 \times 10^{15} p/cm^{2}$ reach a similar capacitance as non-irradiated pixels, at voltages
about 5-10~V higher. For pixels irradiated up to $1 \times 10^{16} p/cm^{2}$, the capacitance remains larger after irradiation, up to the highest voltages measured. Future measurements
will study the performance of sensors optimized for the forward proton timing application, and evaluate the efficiency and time resolution of sensors before and after irradiation.

\section*{Acknowledgments}

This study, and the contribution of C.B.d.C.S., G.D.M., J.H., and M.G., was supported under FCT project PTDC/FIS-PAR/1214/2021. This study has received funding from
the European Union's Horizon Europe Research and Innovation programme under Grant Agreement 101057511. We thank FBK, the RD50 project,  and INFN Torino for providing the LGAD
sensors used in this study. We thank the CERN Solid State Detectors lab for the use of their probe station to perform CV/IV measurements, and for advice on
its operation. We thank the CERN IRRAD group for the efficient irradiation of the sensors, and for helpful discussions on the beamspot profile during the design of the study.

\end{document}